\documentclass{webofc}
\usepackage[varg]{txfonts}   
\usepackage{siunitx}
\usepackage{subcaption}
\usepackage{color}

\begin{document}
\title{Search for Highly Ionizing Particles with the Pixel Detector at Belle II}

\author{\firstname{Katharina} \lastname{Dort}\inst{1}\fnsep\thanks{\email{katharina.dort@physik.uni-giessen.de}} \and
        \firstname{Jens Soeren}  \lastname{Lange}\inst{1}\fnsep\thanks{\email{soeren.lange@exp2.physik.uni-giessen.de}} \and
        \firstname{Klemens} \lastname{Lautenbach}\inst{1}\fnsep\thanks{\email{klemens.lautenbach@physik.uni-giessen.de}}
}

\institute{II Physics Institute, Justus Liebig University Giessen
          }

\abstract{%
	The Belle II experiment, located at the SuperKEKB collider at the high-energy research facility KEK in Tsukuba, Japan, started operation in 2018. Compared to the predecessor experiment Belle, Belle II plans to 
	increase the peak luminosity by a factor of about 40, by employing nano-beam technology in the interaction region.
	In particular the new, innermost sub-detector of Belle II - the Pixel Vertex Detector (PXD) - is in close proximity to the interaction point. This allows for the detection of particles, which do not leave a signal in 
	the outer sub-detectors. Among these, Highly Ionizing Particles (HIPs) encounter a characteristically high energy loss, limiting their penetration depth into the detector.
	Anti-deuterons and magnetic monopoles as possible HIPs are considered. Without a signal in the outer sub-detectors, no track trigger is issued, resulting in possible information loss. The possibility of identifying HIPs solely with information provided by the PXD is presented, by using neural network algorithms operating in a multidimensional parameter space of PXD cluster data. 
}
\maketitle
%

\section{Introduction}

Classical electromagnetism as well as the interaction of charged and electromagnetic
fields are fully described by the Maxwell equations combined with the Lorentz
force law and the Newton equations of motion. The asymmetry of the Maxwell equations due to the existence of electric but the absence of magnetic charges follows no underlying principle other than empirical evidence. The existence of a magnetically charged particle - a so-called magnetic monopole - would symmetrize the Maxwell equation.
\section{SuperKEKB Accelerator and Belle II}

The SuperKEKB accelerator is an asymmetrical electron-positron collider with a center-of-mass energy of 10.58\,GeV~\cite{ohnishi2013accelerator}. As successor of the KEKB accelerator it is located in Tsukuba, Japan. SuperKEKB is designed to reach a world-leading peak luminosity of  $8 \cdot 10^{35} \SI{}{cm^{-2} s^{-1}}$. The interaction point of the collider is surrounded by the Belle II detector~\cite{abe2010belle}. Belle II hosts a silicon-based tracking system consisting of a pixel detector (PXD) and a silicon strip detector (SVD). The PXD consists of DEPFET silicon sensors arranged in two layers. The inner layer has a radial distance of only 1.4\,cm from the interaction point (IP). The close proximity to the IP is ideal for the detection of particles with a short range in matter. The thickness of the silicon sensors is \SI{75}{\micro m} and the pixel size varies along the beam direction from \SI{50 x 50}{\micro m} to  \SI{50 x 85}{\micro m}. In total, it is planned to employ about 8 Million pixels operating with a frame rate of up to 50\,kHz. The entire tracking system is located in a magnetic solenoid field of 1.5\,T. The field lines are aligned with the beam pipe.

The Belle II detector is described by a right-handed Cartesian coordinate system with the origin located at the nominal interaction point and the $z$ axis parallel to the beam pipe. The $y$ axis points to the top of the detector and
the $x$ axis is parallel to the radial direction towards the outside of the detector. In cylindrical coordinates, the polar angle $\theta$ is defined with respect to the $z$ axis and the azimuthal angle $\phi$ is measured in the plane transversal to the $z$ axis.
Additionally, a two dimensional local coordinate system is used for the PXD. The local $v$ axis is parallel to the global $z$ axis and the local $u$ axis to the global $\phi$ axis.

%


\section{Experimental Signature}

 \begin{figure}[tb]
	\centering
	\begin{subfigure}{0.44\textwidth}
		\centering	
		\includegraphics[width=\textwidth]{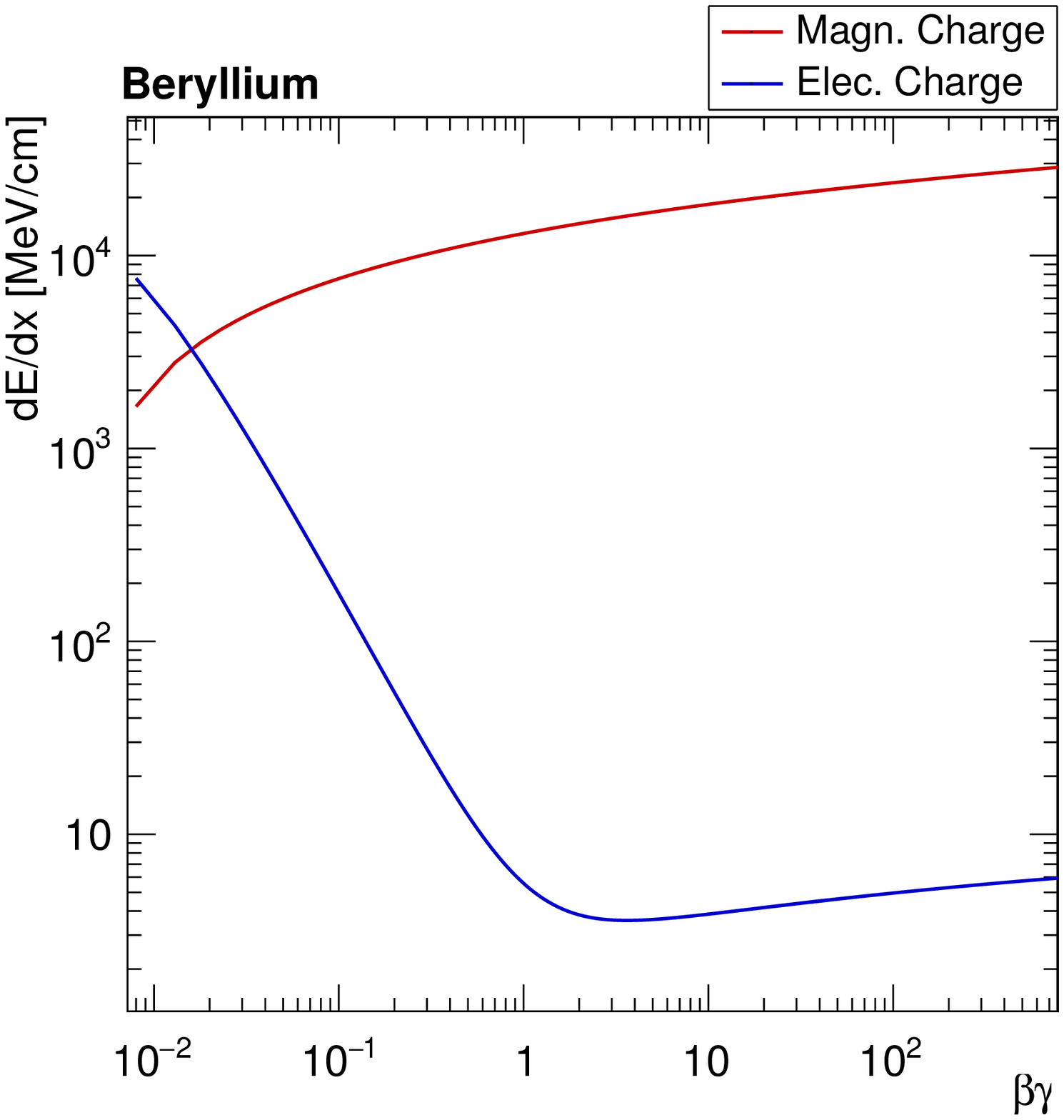}
	\end{subfigure}%
	\begin{subfigure}{0.56\textwidth}
		\centering	
		\includegraphics[width=\textwidth]{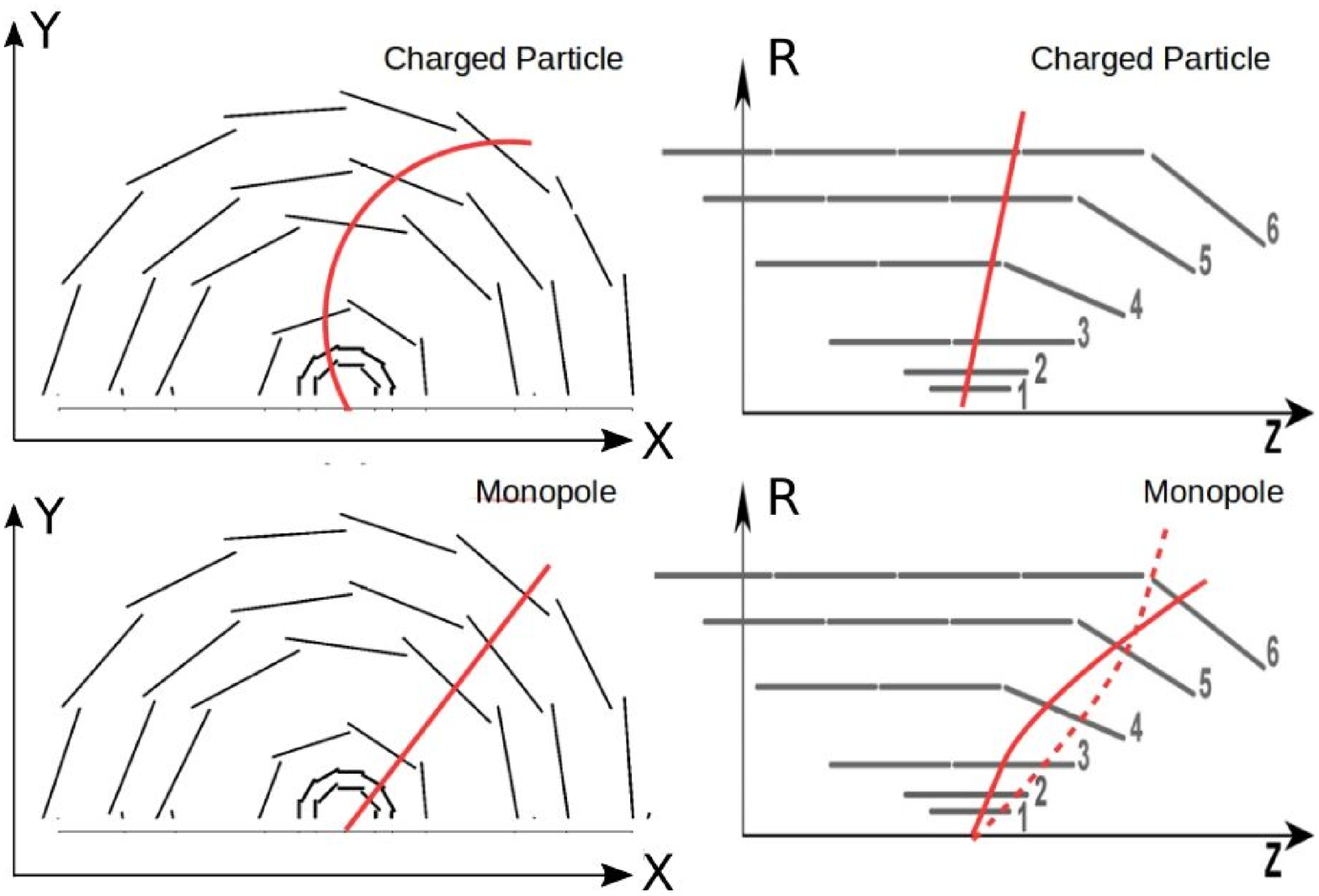}
	\end{subfigure}
	\caption{Energy loss in Beryllium and trajectory of particles with magnetic and electric unit charge. The description of energy loss for magnetic monopoles requires a modification of the Bethe-Bloch curve resulting in an energy loss, which is approximately four orders of magnitude higher compared to electrically charged particles with the same mass. 
	The trajectories of magnetically charged particles in a magnetic field parallel to the $z$ axis differs from the trajectory of electrically charged particles in the same field configuration. }
	\label{fig:mpl_signature}
\end{figure}

Magnetic monopoles appear in various models of particle physics beyond the Standard Model like Grand-Unified Theories~\cite{ross1985grand} and String Theories~\cite{dienes1997string}. 
Our study focuses on magnetic monopoles arising from Dirac's Quantization Theory~\cite{dirac1931quantised}, which predicts the quantization of magnetically charged particles in units of 68.5\,e.  

There are two distinctive features of magnetically charged particles which can be exploited for monopole searches:
\begin{itemize}
	\item \textbf{Energy loss}
	
	The interaction of the monopole's magnetic field with the magnetic moment of atoms leads to an energy loss curve, which differs from the standard Bethe-Bloch formula for electrically charged particle as can be seen in Fig.~\ref{fig:mpl_signature}. As the magnetically charged particle possesses a unit charge of 68.5\,e its energy loss is orders of magnitude higher compared to the particle with unit electric charge 1\,e. 
	The high energy loss limits the range of Dirac magnetic monopoles to the inner vertex region of the Belle II detector. Our study therefore focuses on the identification of magnetic monopoles using only PXD data. In particular, we employ neural networks operating in a mutlidimensional parameter space of PXD data for the detection of magnetically charged particles. We also study the identification of anti-deuterons as a reference signal.
	
	\item \textbf{Trajectory in a magnetic field}
	
	A magnetically charged particle in a magnetic field is accelerated along the magnetic field lines. The monopole track is therefore straight in the $r \phi$ plane and a parabola in the $rz$ plane, when the magnetic field aligns with the $z$ axis as shown in Fig.~\ref{fig:mpl_signature}. The search for unconventional particle tracks at Belle II is not the focus of our approach as it involves fractionally charged monopoles (1\,e, 2\,e, ...) instead of Dirac monopoles (68.5\,e). A separate Belle II study is dedicated to the identification of monopoles via tracking~\cite{conference_neverov}. 
\end{itemize}

\section{Past Searches}

Past searches of magnetic monopoles at electron-positron colliders focused on the identification of monopole tracks in Nuclear Track Detectors (NTDs)~\cite{pinfold1993search,kinoshita1989search,gentile1987search,musset1983search}. Other identification techniques involved tracking~\cite{braunschweig1988search} and the utilization of a wire chamber~\cite{abbiendi2008search}. To our knowledge, the identification of monopoles with silicon detectors has not been attempted yet. All results obtained at a centre-of-mass energy of $\sim$10\,GeV are based upon a data set of $\SI{159}{pb^{-1}}$ of integrated luminosity recorded by the CLEO collaboration~\cite{gentile1987search}. A significant improvement with Belle II data is therefore expected.


\section{Results of Belle II Simulations}
A charged particle impinging on the PXD liberates charge carriers within the silicon sensors of the detector. Typically several adjacent PXD pixels are activated due to the diffusion of charge carriers between the pixel cells. In order to identify a set of pixels with the passage of a single particle they are grouped together to so-called \textit{clusters}. Properties of the cluster such as cluster size (number of activated pixels), cluster charge (total amount of charge carriers registered in the cluster) can be used to identify the particle, which generated the cluster. 

\subsection{Cluster properties}

 \begin{figure}[tb]
	\centering
	\begin{subfigure}{0.49\textwidth}
		\centering	
		\includegraphics[width=\textwidth]{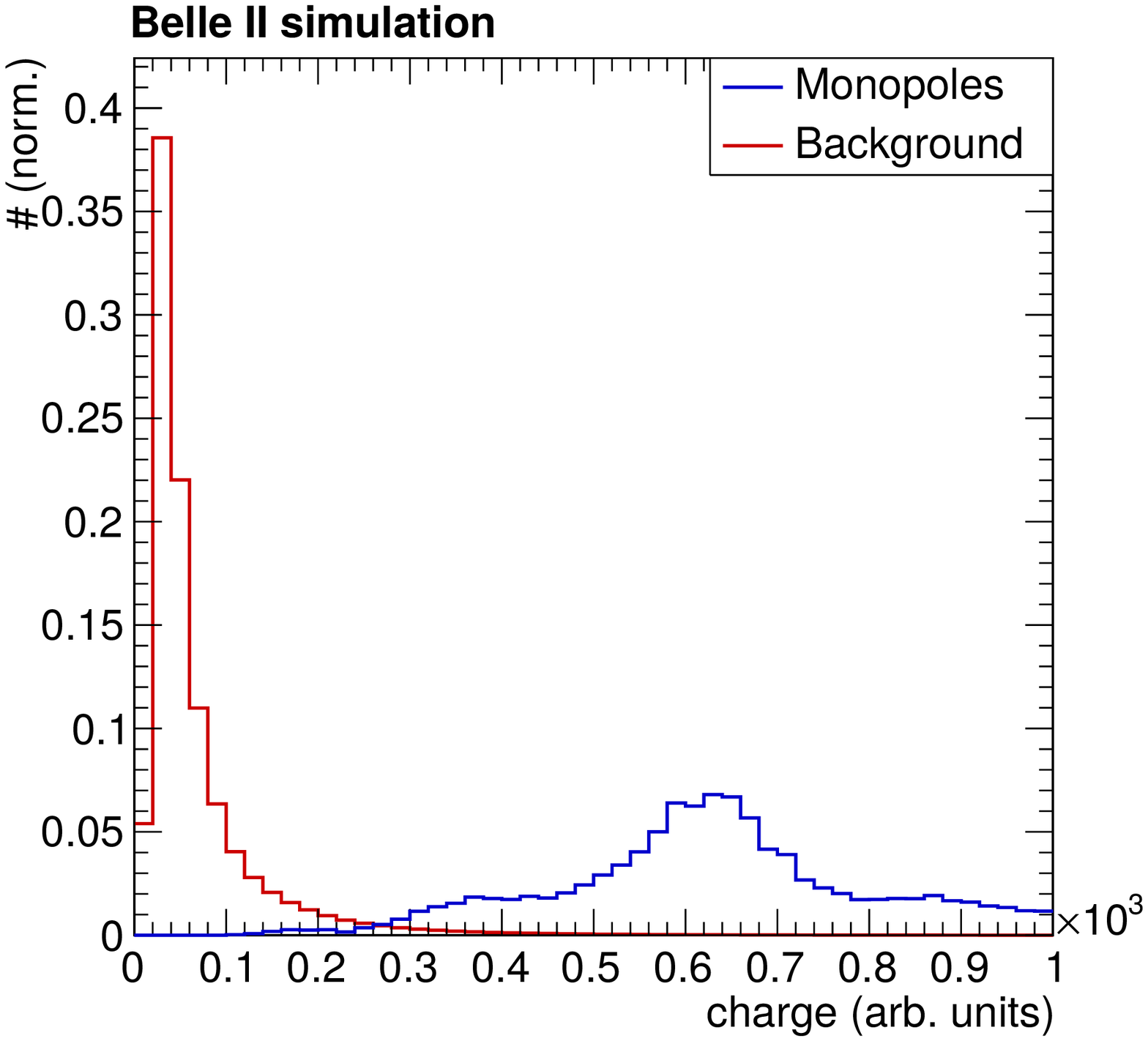}
	\end{subfigure}%
	\begin{subfigure}{0.49\textwidth}
		\centering	
		\includegraphics[width=\textwidth]{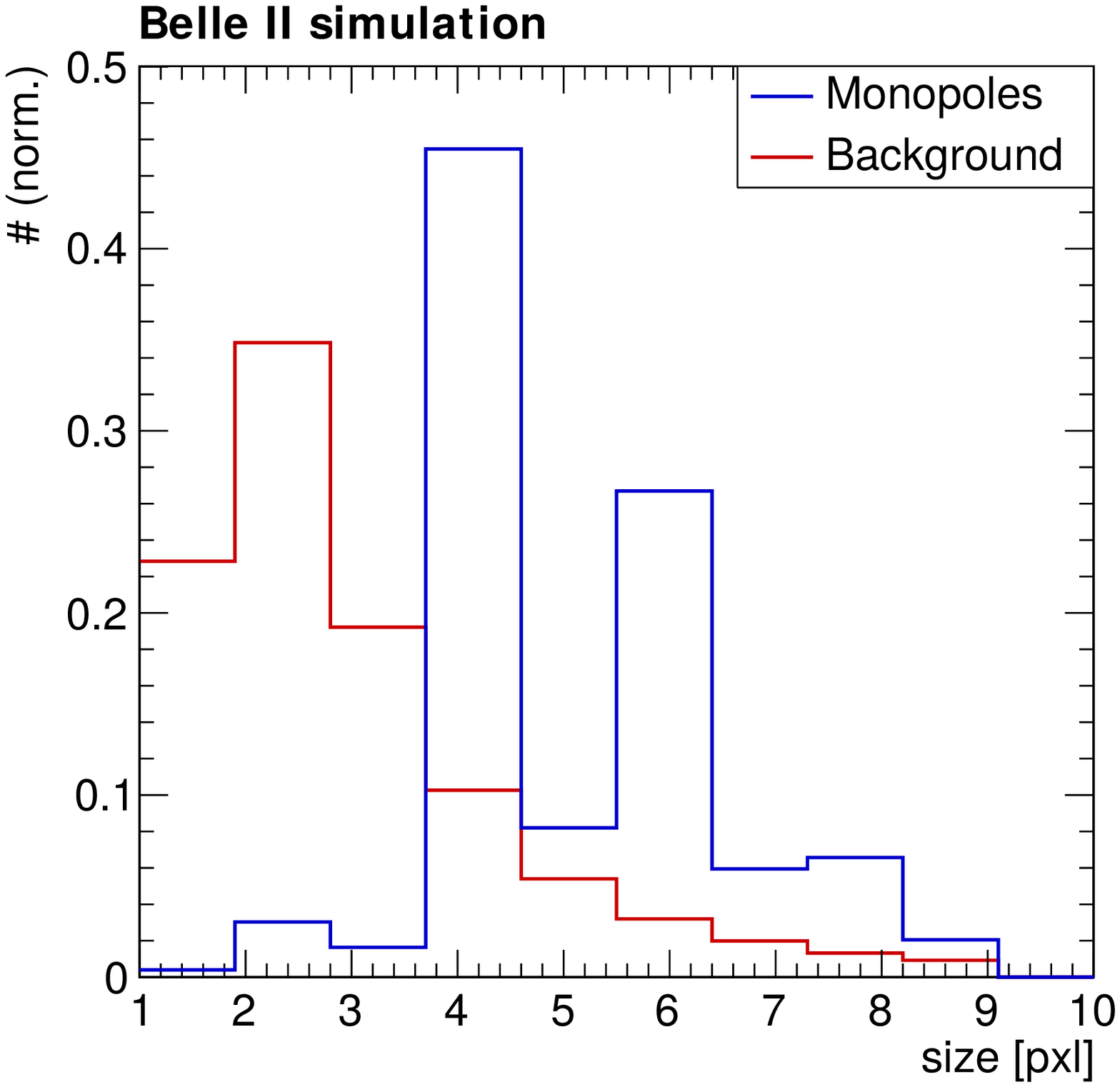}
	\end{subfigure}
	\caption{Cluster charge and cluster size for magnetic monopoles and beam background. Clusters generated by monopoles tend to be significantly higher in charge and size compared to clusters formed by background particles.}
	\label{fig:mpl_clsprp}
\end{figure}

Magnetic monopoles are simulated with the Belle II software framework~\cite{kuhr2019belle}. We focus on the production process $e^+ e^- \rightarrow M \bar M$, where $M$ denotes the magnetic monopole and $\bar M$ an anti-monopole.
The cluster charge and cluster size distributions for magnetic monopoles and beam background are shown in Fig.~\ref{fig:mpl_clsprp}. The monopoles are simulated with unit magnetic charge and a mass of $m = 3$\,GeV. 
On average, clusters generated by magnetic monopoles possess a significantly higher cluster charge and cluster size due to the higher energy deposition of magnetically charged particles in the silicon sensors.
In addition, the acceleration of magnetic monopoles along the magnetic field lines decreases the angle under which the monopoles impinge on the silicon sensors. A shallow angle promotes the creation of charge carriers along several adjacent pixel cells increasing the size of the cluster as well.

\begin{figure}[tb]
	\centering
	\begin{subfigure}{0.49\textwidth}
		\centering	
		\includegraphics[width=\textwidth]{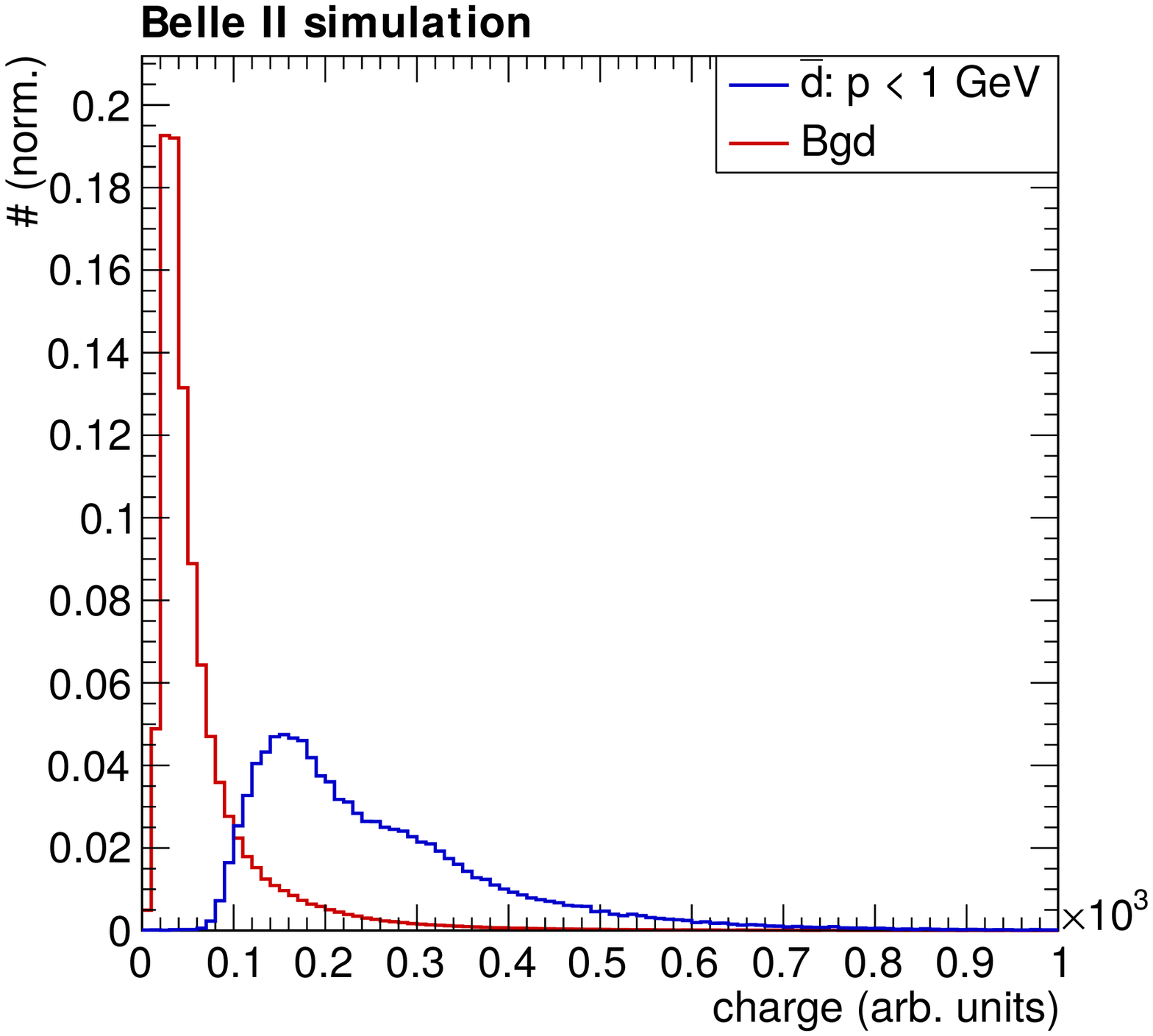}
	\end{subfigure}%
	\begin{subfigure}{0.49\textwidth}
		\centering	
		\includegraphics[width=\textwidth]{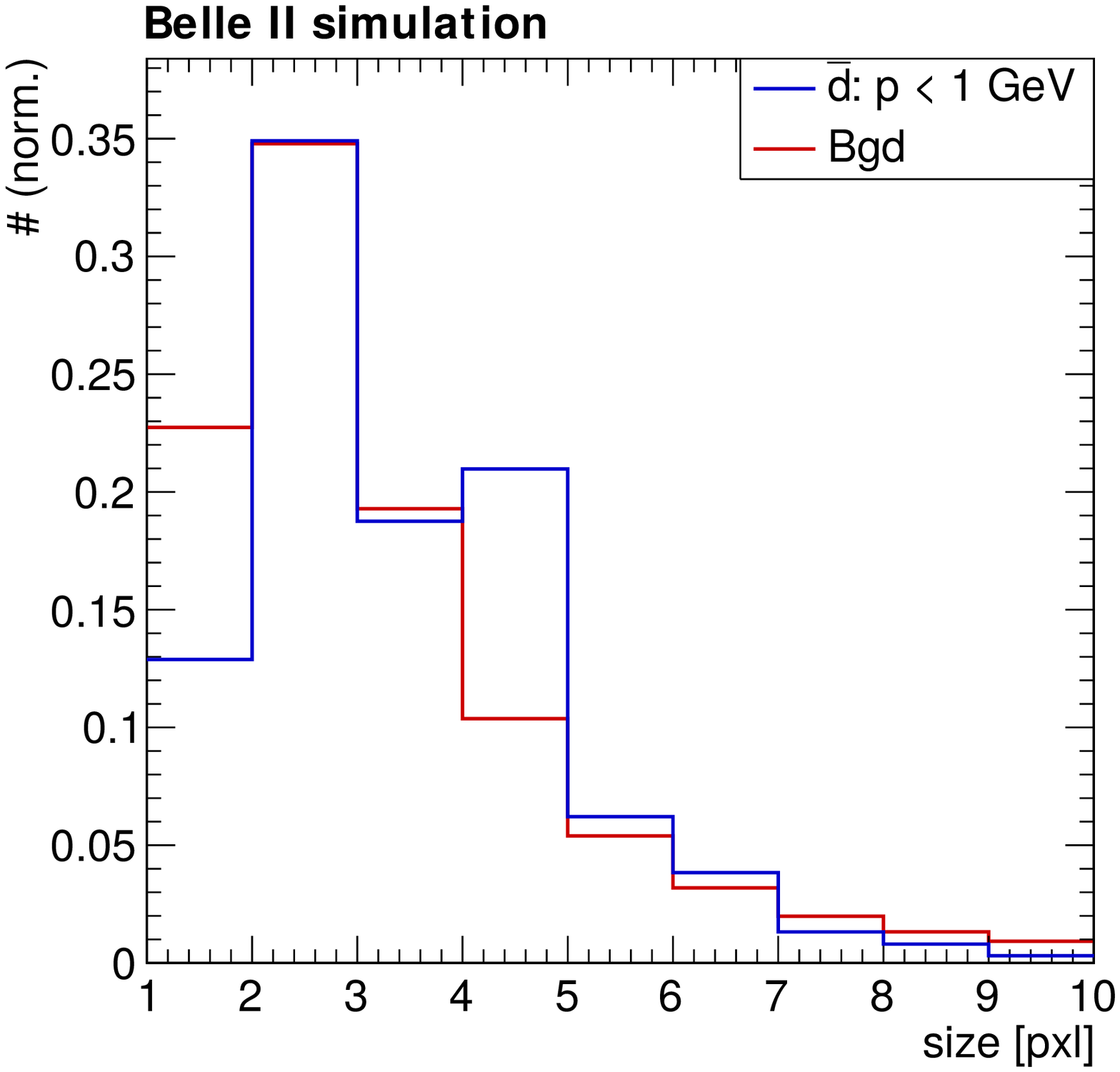}
	\end{subfigure}
	\caption{Cluster charge and cluster size for anti-deuterons and background. 
	Background and anti-deuteron distributions are not separated as neatly as for magnetic monopoles. }
	\label{fig:ad_clsprp}
\end{figure}

The same two distributions for anti-deuterons and background are shown in Fig.~\ref{fig:ad_clsprp}. The overlap of cluster property distributions from anti-deuterons and background is more pronounced compared to magnetic monopoles and background. 

\subsection{Application of Neural Network}

Feed-Forward Neural Networks~\cite{svozil1997introduction} are trained to separate clusters generated by HIPs (magnetic monopoles and anti-deuterons) from clusters generated by background particles. 
The network architecture is presented in Table~\ref{tab:ffn_parameters}. The Feed-Forward Neural Network consists of four fully-connected layers: one input, one output and two hidden layers. Each layer hosts 40 nodes. The activation function is a Rectified Linear Unit (ReLu) and the optimizer employs Stochastic Gradient Descent (SGD). A batch size of 256 has been chosen to comply with the total number of vectors ($\sim $500,000) in each training set. A learning rate of 1E-4 with momentum 0.9 is employed. 

\begin{table}[ht]
	\begin{center}
		\caption{Architecture of Feed Forward Neural Networks used to classify HIPs and background clusters. } 
		\begin{tabular}{|c|c|}
			\hline
			Nr. of layers & 4 \\
			Nodes per layer & 40 \\
			Activation Function & Rectified Linear Unit (ReLu)\\
			Loss Function &  CrossEntropy\\
			Optimizer &  Stochastic Gradient Descent (SGD)\\
			Batch Size &  256\\
			Learning Rate &  1E-4\\
			Momentum &  0.9\\ \hline
		\end{tabular}
		\label{tab:ffn_parameters}
	\end{center}
\end{table}

The input vector set consists of cluster properties (cluster size, size in $u$ direction, size in $v$ direction, cluster charge, seed charge, minimum pixel charge in cluster). 

\begin{figure}[b]
	\centering
	\begin{subfigure}{0.50\textwidth}
		\centering	
		\includegraphics[width=\textwidth]{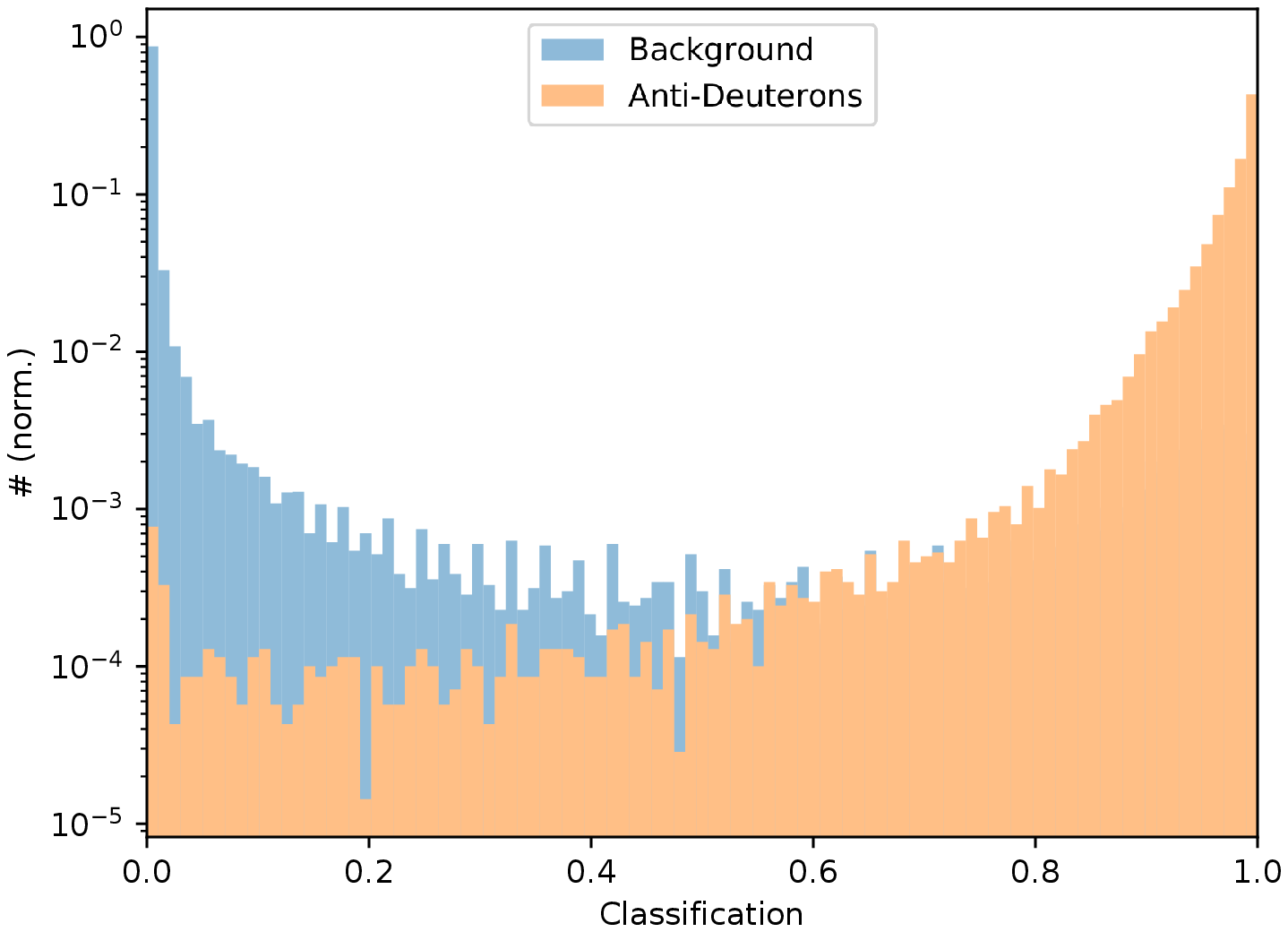}
	\end{subfigure}%
	\begin{subfigure}{0.50\textwidth}
		\centering	
		\includegraphics[width=\textwidth]{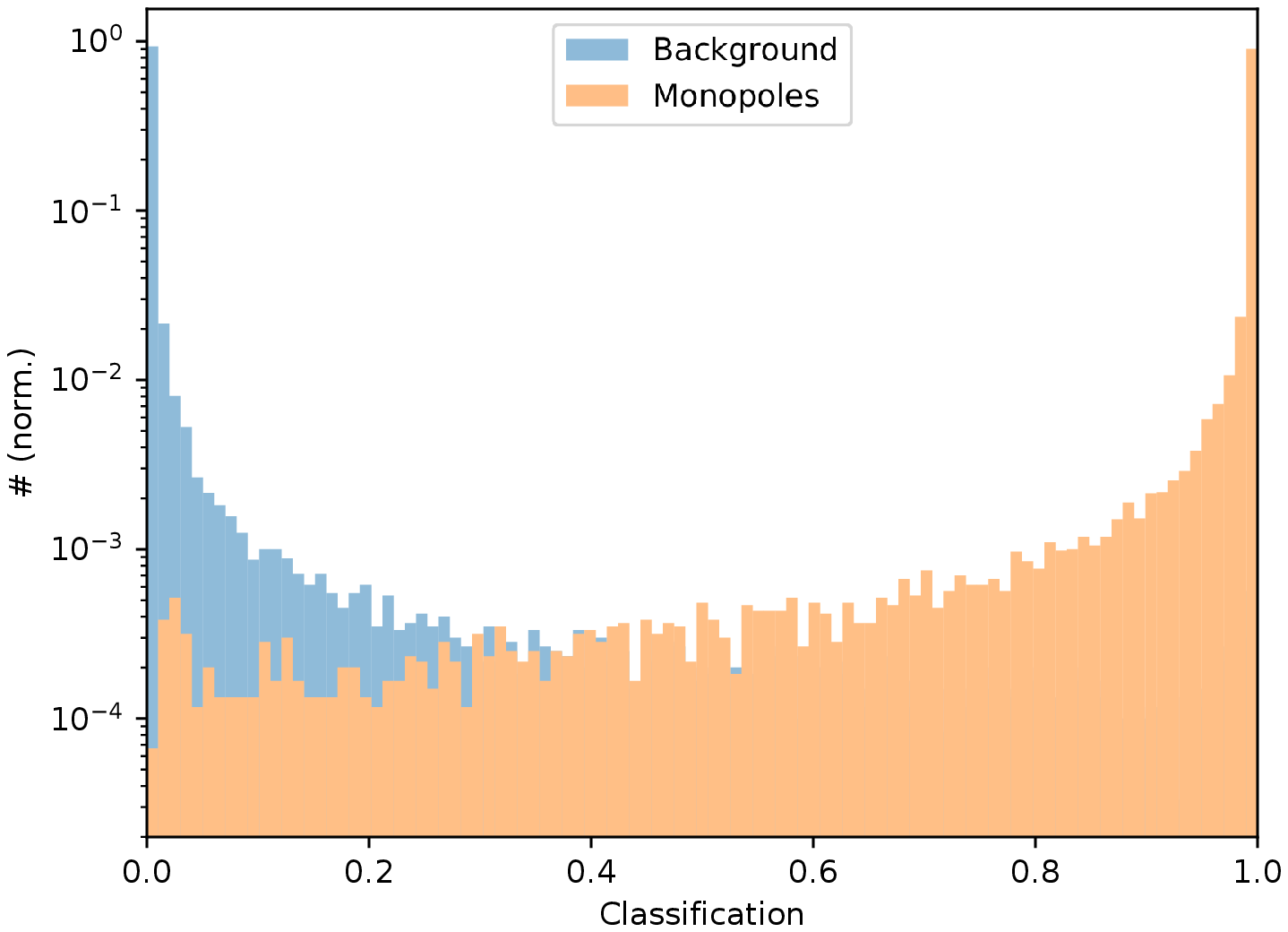}
	\end{subfigure}
	\caption{Classification distribution according to Feed-Forward Network trained with cluster properties of magnetic monopoles and background (right-hand side) as well as anti-deuterons and background (left-hand side). Low classification values correspond to background, high values to signal. The true background is colored in blue, the true signal in orange.  }
	\label{fig:nn_results}
\end{figure}

The classification distributions for Feed-Forward Networks trained with cluster properties of monopoles (right-hand side) and anti-deuterons (left-hand side) against background are shown in Fig.~\ref{fig:nn_results}. Low classification values correspond to background and high values to HIPs. The true HIP distribution is colored in orange, the true background in blue. Background is suppressed by approximately three to four orders of magnitude for high classification values. Likewise, the monopole distribution is several orders of magnitude lower in the low-classification region. A cut on the horizontal axis determines whether an event is classified as background or magnetic monopole. For a cut at 0.5, for instance, all vectors with a classification value greater than 0.5 are associated with monopoles and below 0.5 with background. The cut at 0.5 yields an identification accuracy of $\sim$99\%. 

The separation between anti-deuteron and background clusters is more complicated due to the overlapping cluster property distributions. Nevertheless, the classification distribution demonstrates that a separation of the two cluster types with a neural network is possible. 
A cut on the classification axis at 0.5 allows for an identification accuracy of $\sim$97\%.

\section{Summary and Outlook}
\label{sec:summary}

HIPs can remain undetected in the outer sub-detectors of the Belle II detector due to their short range in the detector material. The identification of HIPs with the innermost sub-detector - the PXD - can prevent a loss of information resulting from the lack of interaction with the outer sub-detectors. The feasibility of separating HIPs from background particles with a neural network has been demonstrated. A detailed study benchmarking the performance of HIP identification is presently underway. 
In the future the identification of HIPs with the PXD could be implemented on FPGAs (Field Programmable Gate Arrays) as part of the PXD readout chain.

\section{Acknowledgment}
This project has received funding from the European Union’s Horizon 2020 research and innovation programme under grant agreement No 644294 as well as the BMBF under grant agreement No 05H15RGKBA and No 05H19RGKBA. 

\bibliography{template.bbl}

\end{document}